\documentclass[aip,
 sd,%
 amsmath,amssymb,
 reprint,%
]{revtex4-1}

\usepackage{graphicx}
\usepackage{dcolumn}
\usepackage{bm}
\usepackage{color}

\begin{document}

\preprint{AIP/123-QED}

\newcommand\eg{\emph{e.g.}\ }
\newcommand\etc{\emph{etc.}}
\newcommand\bcmdtab{\noindent\bgroup\tabcolsep=0pt%
 \begin{tabular}{@{}p{10pc}@{}p{20pc}@{}}}
\newcommand\ecmdtab{\end{tabular}\egroup}
\newcommand\rch[1]{$\longrightarrow\rlap{$#1$}$\hspace{1em}}
\newcommand\lra{\ensuremath{\quad\longrightarrow\quad}}

 \def\be{\begin{equation}}
\def\ee{\end{equation}}

\def\bc{\begin{center}} 
\def\ec{\end{center}}
\def\bea{\begin{eqnarray}}
\def\eea{\end{eqnarray}}
\newcommand{\avg}[1]{\langle{#1}\rangle}
\newcommand{\Avg}[1]{\left\langle{#1}\right\rangle}

\def\adjMatA{{\bf A}}
\def\elemAdjMatA{A}
\def\adjMatB{{\bf B}}
\def\elemAdjMatB{{B}}

\def\dampingA{{\alpha_{A}}}
\def\dampingB{{\alpha_{B}}} 

\def\degreeAIn{k_{\rm in}^{(A)}}
\def\degreeBIn{k_{\rm in}^{(B)}}
\def\degreeBOut{k_{\rm out}^{(B)}}
\def\inDegree{k^{\rm in}}
\def\outDegree{k^{\rm out}}
\def\degreeClass{{\bf k}}
\def\expo{\eta}

\title[Extracting Information from Multiplex Networks]
      {Extracting Information from Multiplex Networks}
      
\author{Jacopo Iacovacci}
 \email{j.iacovacci@qmul.ac.uk}
 \affiliation{School of Mathematical Sciences, Queen Mary University of London, Mile End Road, E1 4NS,  United Kingdom}
 
\author{Ginestra Bianconi}%
 \email{g.bianconi@qmul.ac.uk}
\affiliation{School of Mathematical Sciences, Queen Mary University of London, Mile End Road, E1 4NS, United Kingdom
}
\begin{abstract}
Multiplex networks are generalized network structures that are able to describe  networks in which the same set of nodes are connected by links  that  have different connotations. Multiplex networks are ubiquitous since they describe social, financial, engineering and biological networks as well.  Extending our ability to analyze complex networks to multiplex network structures increases greatly the level of    information that is possible to extract from Big Data. For these reasons characterizing the centrality of nodes in multiplex networks and  finding new ways to solve challenging inference problems defined on multiplex networks are fundamental questions of network science.   In this paper we discuss the relevance of the Multiplex PageRank algorithm for measuring the centrality of nodes in multilayer networks and  we characterize the  utility of the recently introduced indicator function $\widetilde{\Theta}^{S}$  for describing their mesoscale organization and community structure. As working examples for studying these measures we consider three multiplex network datasets coming for social science.
\end{abstract}


\newtheorem{lemma}{Lemma}[section]

\label{firstpage}

\maketitle
\begin{quotation}
Multilayer networks are formed by nodes  connected by links describing interactions with different connotations. When  the same set of nodes can be connected  by different types of links, the resulting multilayer network, is also called a multiplex network. Most social networks are multiplex, since the same set of people might be connected by different types of social ties or might communicate through different means of communication. As networks are ultimately a way to encode information about a complex set of interactions one of the  most pressing  and challenging problem in network science is to extract relevant information from them.  Here we show evidence that the Multiplex PageRank algorithm and the recently introduced indicator function $\widetilde{\Theta}^{S}$ are able to   extract  from multiplex networks an information than cannot be inferred by analyzing  the  single layers taken in isolation or the  aggregated  network in  which links of different type are not distinguished.
\end{quotation}


The vast majority of complex interacting systems, from social networks  to the brain and the biological networks of the cell, are multilayer networks \cite{PhysicsReports,Interdisciplinary,Kivela}.
Multilayer networks  are formed by several  networks (layers) describing interactions of different nature and connotation. 
Therefore multilayer networks encode significant more information than the  network which include all the interactions of the multilayer network but does not distinguishes between the different nature of the links. 
As a consequence of this, one the most pressing challenge in multiplex network theory is devising algorithms and numerical methods to extract relevant information from these network structures.

Multilayer networks can be distinguished in two wide classes: multiplex networks and networks of networks \cite{PhysicsReports,Interdisciplinary}. 
Network of networks are multilayer networks formed by  layers   constituted by different nodes. Examples of network of networks are complex infrastructures, formed for example by interconnected power-grid, Internet and water supply systems and the interacting biological networks of the cell such as  the metabolic network, the protein-protein interaction network, and the transcription network. 

Multiplex networks are another class of multilayer networks. They are formed by the same set of nodes connected by links indicating  different types of interactions. Multiplex networks have been first proposed for modelling social networks \cite{Wasserman}, where the same set of individuals are connected by different types of social ties (friendship, collaboration, family tie, etc.), or can communicate via different means of communication (email, mobile phone, chat, Facebook, etc.). More recently this framework has been used to model a larger set of data including transportation networks \cite{Cardillo,nicosia_correlations}, scientific and actor collaboration networks \cite{menichetti2014weighted,nicosia_correlations}, brain networks \cite{Bullmore,Makse,Battiston_measures}, and biological networks in the cell \cite{Caselle}. The increasing interest in multiplex network has also revealed that the multiplex network structure strongly affects the behavior of dynamical processes \cite{PhysicsReports, Kivela,Cellai2013,Bianconi2014,Cellai2016,synchronization1,synchronization2,epidemic,gametheory1,gametheory2,congestion, Buldyrev2010,Goh_rev,Diaz,Radicchi}.

Multiplex networks display usually  a   correlated network structure. In fact they are very often characterized by a  significant overlap of the links in different layers \cite{Cardillo,szell2010multirelational,PRE,menichetti2014weighted},  by correlations between the degree of the same node in different layers \cite{Goh_correlations,Battiston_measures},  by the heterogeneous activity (presence of a node in a layer) of the nodes in different layers \cite{nicosia_correlations, Cellai_activity} and by a significant overlap of the communities in different layers \cite{iacovacci2015,battiston2015emergence}.
These correlations can be exploited to extract relevant information from multiplex network structures that cannot be inferred by analyzing single layers taken in isolation or the aggregated network where all the interactions are taken at the same level.

In this paper our aim is to extract relevant information from multiplex networks datasets describing social interactions. In particular we will focus on  how to 
characterize the centrality of the nodes in multiplex networks and how to  characterize the mesoscale organization of multiplex datasets, providing three major case studies of multiplex social networks where the proposed algorithms are applied.

Quantifying the centrality of the nodes in multiplex networks is a fundamental problem that in these last years has attracted  increasing attention from the network science community \cite{sola2013eigenvector,MultiplexPageRank,de2015ranking}. Among the different proposed centrality measures, the Multiplex PageRank \cite{MultiplexPageRank} can be used to assess how the centrality of a node in a single layer is affected by the centrality of the same node in another layer. This effect is significant for example in social networks where the reputation of an individual in a layer can affect its centrality in another layer. Here we will show how the Multiplex Page Rank apply to the { Physical Review E} Citation-Collaboration Multiplex Network \cite{menichetti2014weighted} of scientists that collaborate with each other and cite each other. In this context, if a  scientist collaborates typically  in small teams, his centrality  in the citation network can boost his centrality and the centrality of his neighbors in the collaboration network because these collaborations are more prestigious. Also the reverse can happen, and  the centrality of a scientist in the collaboration network can boost his centrality and the centrality of his neighbors in the citation network because he is playing the role of a  catalyst of scientific activity. 

{ Moreover we apply the Multiplex PageRank to the fully annotated and public Noordin Multiplex Terrorist Network \cite{nancy2011}, where relations between the people inside a terrorist organization are described at three different levels: the level of the trust, the level of the communication and at the level of the co-operation. For this reason the Noordin Terrorist Network dataset represents an interesting case study to perform the Multiplex PageRank algorithm extended to more than two layers. Indeed, assuming that the prominence of a node in the layer of the co-operation is affected by its prominence in the layer of the communication, and the prominence in the layer of communication is itself affected by the prominence of the node in the layer of the trust, we will show that the Multiplex PageRank is able to assess the centrality and the role that the terrorists play in their organization.}         

In a multiplex network additional relevant information is encoded in the mesoscale structure of the layers. In particular, in a large variety of cases the communities in one layer have a very significant overlap with the communities of other layers. This property  is  routed in the way these networks evolve \cite{battiston2015emergence} and reveal an important mechanism by which  multiplex networks encode information about their function. 
Several works address the problem of detecting the community structure of multiplex networks \cite{Mucha,Sales,de2014identifying, MuxViz,iacovacci2015}. 
Recently a new approach \cite{iacovacci2015} has been proposed to solve the relevant problem of qualifying the similarities between the mesoscale structure of  the layers of  multiplex networks. This  new framework is able to extract for any multiplex, the network between its layers weighted by a similarity score of their  mesoscale structure.  Here we will show how this algorithm applies to the  CS-Aarhus Collaboration Network \cite{magnani2013}  formed by 5 layers of online and offline relationships (Facebook, Leisure, Work, Co-authorship, Lunch) between the 61 employees of Computer Science department at Aarhus.

The paper is organized as follows. In Sec.$\ref{ii}$ we will define the Structural Multiplex Measures considered in this paper including the Multiplex PageRank and the Indicator function of similarity between the multiplex layers. In Sec. $\ref{iii}$ we will show how these measures apply to three social multiplex network datasets including the { Physical Review E} Citation-Collaboration Multiplex Network, The Noordin { Multiplex} Terrorist Network and the CS-Aarhus { Multiplex} Collaboration Network.
Finally in Sec. $\ref{iv}$ we will give the conclusions.

\section{Structural Multiplex Measures}
\label{ii}
\subsection{Multiplex Networks}

A multiplex network is formed by $N$ nodes $i=1,2\ldots, N$ and $M$ layers $\alpha=1,2,\ldots,M$. Each layer is a network described by a given  adjacency matrix ${\bf a}^{\alpha}$. The layers of a multiplex network can be weighted or unweighted, directed or undirected. If the  layers are formed by simple networks (unweighted and undirected) each  element of any of the adjacency matrices takes the value $a_{ij}^{\alpha}=1$ if node $i$ is connected to node $j$ in layer $\alpha$, and is set to zero otherwise, i.e. $a_{ij}^{\alpha}=0$. 
The degree of a node  $i$ on layer  $\alpha$ is indicated with  
\bea
k_{i}^{\alpha}=\sum_{j=1}^N a_{ij}^{\alpha}.
\eea 
For weighted undirected  layers each  element of any of the  adjacency matrices takes the value $a_{ij}^{\alpha}=w_{ij}^{\alpha}$ if node $i$ is connected to node $j$  in layer $\alpha$ by a link of weight $w_{ij}^{\alpha}$ , and takes the value $a_{ij}^{\alpha}=0$ otherwise. 
In this case the degree $k_i^{\alpha}$  of node $i$ in layer $\alpha$  is  given by 
\bea
k_{i}^{\alpha}&=&\sum_{j=1}^N \theta\left(a_{ij}^{\alpha}\right),
\eea
where $\theta(x)=1$ if $x>0$ and $\theta(x)=0$ if $x\leq0$.
For directed unweighted layers the  elements of the adjacency matrix take the value $a_{ij}^{\alpha}=1$ if node $j$ points to node $i$ in layer $\alpha$, and $a_{ij}^{\alpha}=0$ otherwise. 
In this case we distinguish between the in-degree $k_i^{in,\alpha}$ and the out-degree $k_i^{out,\alpha}$ of node $i$ given by 
\bea
\begin{array}{lllrll}
k_i^{in,\alpha}&=&\sum_{j=1}^N a_{ij}^{\alpha},& k_i^{out,\alpha}&=&\sum_{j=1}^N a_{ji}^{\alpha}.
\end{array}
\eea
Finally for  directed weighted layers the  elements of the adjacency matrix take the value $a_{ij}^{\alpha}=w_{ij}^{\alpha}$ if the  directed link from node $j$ to node $i$ has weight $w_{ij}^{\alpha}$, and $a_{ij}^{\alpha}=0$ otherwise.
In this case we distinguish between the in-degree $k_i^{in,\alpha}$ and  the out-degree $k_i^{out,\alpha}$ of node $i$ in layer $\alpha$ given by 
\bea
\begin{array}{lllrll}
k_i^{in,\alpha}&=&\sum_{j=1}^N \theta\left(a_{ij}^{\alpha}\right), & k_i^{out,\alpha}&=&\sum_{j=1}^N \theta\left(a_{ji}^{\alpha}\right).
\end{array}
\eea
Multiplex networks are characterized by correlations of different nature \cite{PhysicsReports,PRE,Goh_correlations,nicosia_correlations}. One way to measure correlations  between the layers is to calculate the correlations between the  degree that each node has in different layers, which can be either  positively correlated or anti-correlated. 
Multiplex networks might also display a significant overlap of the links, indicating that the number of links common to different layers is large compared to a null hypothesis where the links are randomly distributed among the nodes. Finally also the community structure of networks in different layer might be correlated, with communities defined in different layers overlapping with each other.
Inference problems on multiplex networks, including the determination of centrality measures for nodes \cite{sola2013eigenvector,MultiplexPageRank,de2015ranking} and the characterization of the multiplex network mesoscale structure \cite{Mucha,Sales,de2014identifying, MuxViz,iacovacci2015,battiston2015emergence}, can take advantage of these correlations to extract more information from these datasets. In this way a clear path is defined for extracting relevant information from multiplex network structures, which cannot be unveiled by analyzing  its single layers taken in isolation or  its aggregated description.

In the following we will show that the recently introduced measures of the Multiplex PageRank \cite{MultiplexPageRank} and the indicator function $\widetilde{\Theta}^S$ \cite{iacovacci2015} are valuable measures for assessing the centrality of nodes in multilayer networks and for characterizing the mesoscopic similarity between the layers of these structures.

\subsection{Multiplex PageRank}

The Multiplex PageRank \cite{MultiplexPageRank} evaluates the centrality of the nodes  of  multiplex networks. The main effect that the Multiplex PageRank aims at capturing is the influence of the centrality of a node in one layer to its centrality in another layer. Assume for example that we consider  a very central  actor in the movie collaboration network. If the famous actor takes part in social causes its centrality in the actor movie collaboration network might influence its centrality in socio-political causes. This is the case for example with famous actors such as Angelina Jolie that is also a UN Goodwill ambassador. 
Therefore, in  the Multiplex PageRank the centrality of a node in one layer might effect the centrality of the same node in other layers.

In order to capture this phenomena we choose a master layer $\alpha$ with adjacency matrix $a^{\alpha}_{ij}=A_{ij}$ and we calculate the PageRank $x_i$ of a node $i$ in a network \cite{brin1998anatomy} 
\begin{equation}
x_i = \mu_A \sum\limits_{j} \elemAdjMatA_{ij} \frac{x_j}{g_j} + \theta_A \frac{1}{N},
\label{Pagerank}
\end{equation}
where $g_j=\tilde{g}_j+\delta(\sum_{r}\elemAdjMatA_{rj},0)$, with  $\tilde{g_j}=\sum_r A_{rj}$, and  $\delta(x,y)$ indicating the Kroneker delta, i.e. $\delta(x,y)=1$ if $x=y$ otherwise and $\delta(x,y)=0$. Here  $\mu_A>0$ indicates the damping factor and $\theta$ indicating the teleportation parameter, given by 
\begin{equation}
\theta_A=\sum_j \left[1-\mu_A+\mu_A\delta(\tilde{g}_j,0)\right]x_j.
\end{equation}
Usually the PageRank is calculated on directed unweighted networks but the definition of PageRank can also be extended to weighted networks. 
If the network is unweighted,  the PageRank represents the stationary distribution of a random walker that can either hop from a node to a neighboring node or perform a random jump to any node of the network. A random walker on a site $j$ with at least out-degree one, hops to one of $j$'s $\outDegree_j$ out-neighbors with probability $\mu_A$, and to any other site chosen uniformly at random with probability $1-\mu_A$. 
If, instead, the   adjacency matrix $A_{ij}$ is weighted, the  random walker on site $j$ with at least out-degree one hops to one of $j$'s $\outDegree_j$ out-neighbors with a probability depending on the weight of the link between node $j$ and node $i$, and  given by   $\mu_A A_{ij}/g_j$ or performs a random jump to an arbitrary node with probability $1-\mu_A$.\\
 In both cases, if the random walker is on a site $j$ with zero out-degree, it jumps to random node with probability one.

The damping factor $\mu_A$ should be taken smaller than the inverse of the maximum eigenvalue $\lambda_{max}$ of the matrix $C$ with matrix elements $c_{ij}=\elemAdjMatA_{ij}/{g_j}$. For undirected networks this upper bound is one, for directed unweighted networks, one empirical value that is often taken is $\mu_A=0.85.$

Given the rank $x_i$ of node $i$ in the layer $\alpha$, the Multiplex PageRank algorithm measures that the centrality of this node in any other layer $\alpha'$ (called also layer B) with adjacency matrix $a_{ij}^{\alpha'}=B_{ij}$ using a random walker, as in the usual PageRank. Nevertheless, the random walker of the Multiplex PageRank has a probability of visiting the nodes that is affected by their rank $x_i$ in layer $\alpha$. {In other words, the Multilayer PageRank evaluates  the rank of the nodes in layer B by  determing the stationary probability of   a random walk in layer B, biased by the PageRank of the nodes in layer A.}

We distinguish  the following three cases of   Multiplex PageRank :
\begin{itemize}
\item {\em Additive Multiplex PageRank}

The Additive Multiplex Page Rank $X_i^{[ADD]}$ of node $i$ in layer B, is determined by the recursive equation
\begin{equation}
X_i^{[ADD]} = \mu_B \sum\limits_j B_{ij} \frac{X_j^{[ADD]}}{G_j} + \theta_B\frac{x_i}{N\avg{x}},
\label{additive}
\end{equation}
where $G_j=\tilde{G}_j+\delta(\sum_{r}B_{rj},0)$, $\tilde{G}_j=\sum_{r}B_{rj}$ and 
\begin{equation}
\theta_B=\sum_j \left[1-\mu_B+ \mu_B\delta(\tilde{G}_j,0)\right]X_j^{[ADD]}.
\end{equation}
In this case the random walker,  initially  on a node $j$  with out-degree at least one, with probability $\mu_B$ hops to  a node $i$, chosen among the outgoing neighbors of the node $j$  while with probability $(1-\mu_B)$ it  jumps to a random node $i$  chosen  according to this centrality $x_i$ in layer A. If the random walker on node $j$ hops to a  neighbor node $i$, this node  is chosen among the other neighbors of node $j$ with  probability $B_{ij}/G_j$ dictated exclusively by the weights of the interaction in layer B. If instead the random node is on a node $j$ that does not have outgoing neighbors, the random walk jumps to a random node of the network chosen according to this centrality $x_i$ in layer A. 

Here each node in layer  B derives an added benefit  by being central in network A, regardless of the relevance of the nodes that point to it in layer B.

\item{\em Multiplicative Multiplex PageRank }:

The Multiplicative Multiplex Page Rank $X_i^{[MUL]}$ of node $i$ in layer B, is determined by the recursive equation
\begin{equation}
X_i^{[MUL]} = \mu_B \sum\limits_j x_i B_{ij} \frac{X_j^{[MUL]}}{G_j} + \theta_B\frac{1}{N},
\label{multiplicative}
\end{equation}
where $G_j=\tilde{G}_j+\delta(\sum_{r}B_{rj},0)$, $\tilde{G}_j=\sum_{r}B_{rj}$ and 
\begin{equation}
\theta_B=\sum_j \left[1-\mu_B +\mu_B\delta(\tilde{G}_j,0)\right]X_j^{[MUL]}.
\end{equation}
In this case the random walker, initially  on a node $j$  with out-degree at least one, with probability $\mu_B$ hops to  a node $i$, chosen among the outgoing neighbors of the node $j$  while with probability $(1-\mu_B)$ it  jumps to a random node $i$  chosen  with uniform probability. If the random walker on node $j$ hops to a  neighbor node $i$, this node  is chosen among the other neighbors of node $j$ with probability proportional to $x_i B_{ij}/G_j$, i.e. proportionally to the PageRank of node $i$ in layer A. If instead the random node is on a node $j$ that does not have outgoing neighbors, the random walk jumps to a random node of the network chosen with uniform probability. 
Here  all benefits from being central in network A are contingent upon the connections that a node receives from central  nodes in network B.

\item{\em Combined Multiplex PageRank }:
The Combined Multiplex Page Rank $X_i^{[COM]}$ of node $i$ in layer B, is determined by the recursive equation
\begin{equation}
X_i^{[COM]} = \mu_B \sum\limits_j x_i B_{ij} \frac{X_j^{[COM]} }{G_j} + \theta_B\frac{x_i}{N\avg{x}},
\label{combined}
\end{equation}
where $G_j=\tilde{G}_j+\delta(\sum_{r}B_{rj},0)$, $\tilde{G}_j=\sum_{r}B_{rj}$ and 
and 
\begin{equation}
\theta_B=\sum_j \left[1-\mu_B+\mu_B\delta(\tilde{G}_j,0)\right]X_j^{[COM]}.
\end{equation}
The random walker, initially  on a node $j$  with out-degree at least one, with probability $\mu_B$ hops to  a node $i$, chosen among the outgoing neighbors of the node $j$  while with probability $(1-\mu_B)$ it  jumps to a random node $i$  chosen  according to this centrality $x_i$ in layer A. If the random walker on node $j$ hops to a  neighbor node $i$, this node  is chosen among the other neighbors of node $j$ with  probability $x_i B_{ij}/G_j$, i.e. proportional to its ranking $x_i$ in layer A. If instead the random node is on a node $j$ that does not have outgoing neighbors, the random walk jumps to a random a random node   of the network chosen  according to this centrality $x_i$ in layer A.
The effect of network A on network B is a combination of the effects of an additive and of a multiplicative PageRank.
\end{itemize}
In all three cases the damping factor $\mu_B$ should be greater than zero, and smaller that the maximal eigenvalue of the hopping matrix $C$ with elements $c_{ij}$ taking values 
$c_{ij}=B_{ij}/{G_j}$ for the Additive Multiplex PageRank and $c_{ij}=x_i B_{ij}/{G_j} $ for the Multiplicative and Combined Multiplex PageRank.

The Multiplex PageRank can be applied repeatedly to different layers of a given Multiplex Network, as will be shown in Sec. $\ref{noordin}$. 

\subsection{Assessing the similarity between the layers of a Multiplex Network }
\label{thetasec}
The indicator function $\widetilde{\Theta}^S_{\alpha,\beta}$ \cite{iacovacci2015} characterizes the similarity between any two  layers $\alpha, \beta$ of a multiplex network, based on their mesoscale structure.
When the full matrix $\widetilde{\Theta}^S_{\alpha,\beta}$  $\forall \alpha,\beta$ is given, it is possible to construct a network formed by nodes representing  the layers of the multiplex network and by links indicating their similarity. This allows to have a description of the multilayer network in terms of a network of networks between its layers.

The indicator function $\widetilde{\Theta}^S_{\alpha,\beta}$ is based on the entropy of network ensembles \cite{bianconi2008entropy,bianconi2009entropy,peixoto2012entropy,bianconi2009assessing}.
A given network  ensemble  is the set of all networks satisfying a number of structural constraints.
The entropy of a network ensemble indicates the logarithm of the typical number of networks in the ensemble. This quantity is an information theory measure to evaluate the information level encoded in the imposed constraints. In fact if the constraints are very demanding and difficult to satisfy, we expect that the number of networks satisfying these constraints will be rather small. This case will yield a small entropy of the network ensemble and the  large level of information in the constraints. On the contrary,  if the constraints are very easy to satisfy the number of networks in the ensemble will be large. In this case the entropy will be large and the constraint will carry a smaller level of information.

Given a specific multiplex network we can compare it with  null models for the real dataset generated by suitable ensemble of multiplex networks. 

To this end we identify the characteristics $q_{i}^{\alpha} \in \{1,\ldots, Q^{\alpha}\}$, associated to each node $i$ of layer $\alpha$. The characteristics $q_i^{\alpha}$ can  indicate the community assignment of node $i$ in layer $\alpha$ but can also indicate  other features of the node $i$ in layer $\alpha$.
Nodes can  thus be  distinguished in $P^{\alpha}$ classes $p_i^{\alpha}\in \{1,\ldots, P^{\alpha}\}$ which are function of the degree of the nodes and of their characteristic $q_i^{\alpha}$, i.e. $p_i^{\alpha}=f(k_i^{\alpha},q_i^{\alpha})$. For example, the function $f$ can be taken to take different values if for each pair of $(k_i^{\alpha},q_i^{\alpha})$. Assuming that the classes of the nodes depend both on their degree and their characteristic is a  minimal assumption able to describe  networks with communities and strong degree heterogeneities. 
Using the assignment $p_i^{\alpha}$ of nodes into classes, we can extract from the multiplex network under study its block structure. This is described by the  the matrices $\textbf{e}^{\alpha}$ of elements $e^{\alpha}(p,p')$  indicating the total number of links on the layer $\alpha$ between nodes of class $p$ and nodes of class $p'$. 
Finally we can construct a null model of the given multiplex networks by  considering the multiplex network ensemble  with the same block structure $e^{\alpha}(p,p')$ of the given network dataset. 

The entropy $\Sigma_{k^{\alpha},q^{\alpha}}$ \cite{bianconi2008entropy,bianconi2009entropy,peixoto2012entropy,bianconi2009assessing} of a generic layer $\alpha$ is given by 
\begin{equation}
\Sigma_{k^{\alpha},q^{\alpha}}=\log\left[\prod_{p<p'}\left(\begin{array}{c} n_p^{\alpha}n_{p'}^{\alpha} \label{entropy}\\ e^{\alpha}(p,p')\end{array}\right)\prod_p \left(
\begin{array}{c} n_p^{\alpha}(n_{p}^{\alpha}-1)/2 \\ e^{\alpha}(p,p)\end{array}\right)\right]
\end{equation}
where $n_p^{\alpha}$ is the number of nodes having class $p$ in layer $\alpha$ and $e^{\alpha}(p,p')$ is the number of links between nodes having class $p$ and nodes having class $p'$ on layer $\alpha$. 

The entropy  $\Sigma_{k^{\alpha},q^{\alpha}}$ measures how much information is encoded in the block structure imposed to the network. If given the assignment $\{q^{\alpha}_i\}$ the entropy is much smaller than in a random hypothesis (when the characteristics are reshuffled randomly between the nodes), this means that the characteristics $\{q^{\alpha}_i\}$ capture relevant information respect to the network structure $\{k^{\alpha}_i\}$. We can thus quantify the specificity of a generic layer $\alpha$ respect to the assignment $q_i^{\alpha}$ using the Z-score function $\Theta$ proposed in \cite{bianconi2009assessing}:  
\begin{equation}
\Theta_{k^{\alpha},q^{\alpha}}=\frac{E_{\pi}[\Sigma_{k^{\alpha},\pi(q^{\alpha})}]-\Sigma_{k^{\alpha},q^{\alpha}}}{\sigma_{\pi}[\Sigma_{k^{\alpha},\pi(q^{\alpha})}]}
\end{equation}
where $E_{\pi}[...]$ is the expected value over random uniform permutations $\pi(q^{\alpha})$ of the node characteristics $q^{\alpha}$ in layer $\alpha$ and $\sigma_{\pi}[...]$ the corresponding standard deviation.

This function can be used to compare the similarity between the different layers of a multiplex network. 
We first measure how much information the characteristics $q^{\beta}$ contain respect to the node structure of layer $\alpha$ by the corresponding indicator $\Theta_{k^{\alpha},q^{\beta}}$:
  
\begin{equation}
\Theta_{k^{\alpha},q^{\beta}}=\frac{E_{\pi}\left[\Sigma_{k^{\alpha},\pi(q^{\beta})}\right]-\Sigma_{k^{\alpha},q^{\beta}}}{\sigma_{\pi}\left[\Sigma_{k^{\alpha},\pi(q^{\beta})}\right]}.
\end{equation}

In order to assess the significance of the obtained values of $\Theta_{k^{\alpha},q^{\beta}}$, we normalize this quantity by value obtained by considering the value of $\Theta$ induced in layer $\alpha$ by its block structure $q^{\alpha}$. We define therefore the quantity
\begin{equation}
\widetilde{\Theta}_{\alpha,\beta}= \frac{\Theta_{k^{\alpha},q^{\beta}}}{\Theta_{k^{\alpha},q^{\alpha}}}.
\end{equation}
 The quantity $\widetilde{\Theta}_{\alpha,\beta}$ is a measure of how the structure of layer $\beta$ is similar to the structure of layer$\alpha$ respect to the community assignment. In particular  $\widetilde{\Theta}_{\alpha,\beta}=1$ when the community structure $q^{\beta}$, proper of layer $\beta$, carries the same level of information for the structure of layer $\alpha$ as the community structure $q^{\alpha}$, proper of layer $\alpha$. 
 
 \begin{figure}
\includegraphics[width=0.9\columnwidth]{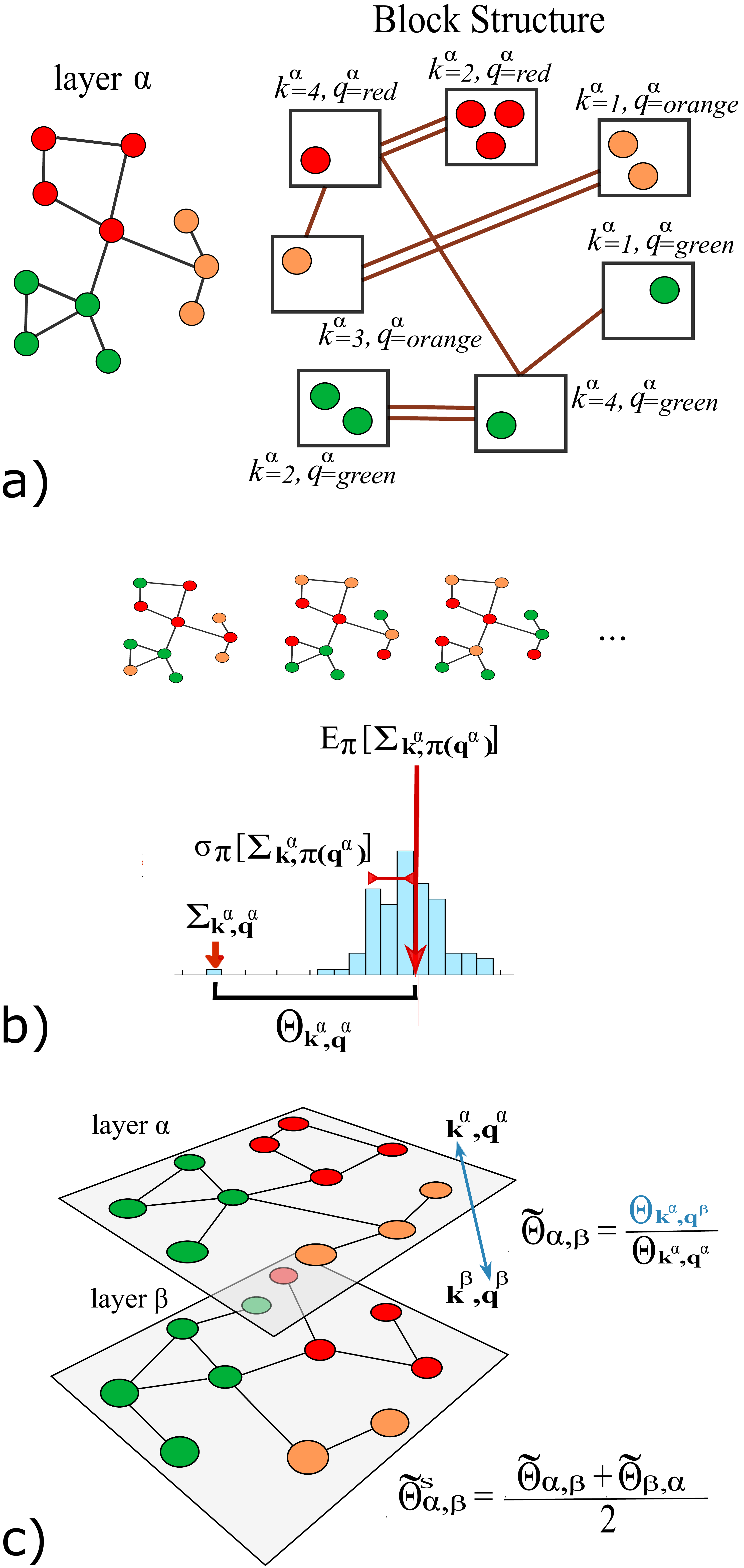} 
  \caption{  Schematic representation  of  the method used to calculate $\widetilde{\Theta}_{\alpha,\beta}^{S}$. Panel a): Given a layer $\alpha$ of a multiplex network,  the node classes  $p^{\alpha}=(k^{\alpha},q^{\alpha})$ are defined, where $k^{\alpha}$ indicates the node degrees  
  and $q^{\alpha}$ the node characteristics on the layer $\alpha$. {These classes induce a block structure in the network specified  by  the number of links between the nodes of each class and the number of links connecting the nodes in different classes}. Panel b):  The entropy $\Sigma_{k^{\alpha},q^{\alpha}}$ given by Eq. $(\ref{entropy})$ is calculated and compared with the entropy distribution obtained in a random hypothesis, by performing random uniform permutations $\pi(q^{\alpha})$ of the characteristics $q^{\alpha}$ of the nodes and subsequently measuring the $\Sigma_{k^{\alpha},\pi(q^{\alpha})}$ values. The mean $ E_{\pi}\left[\Sigma_{k^{\alpha},\pi(q^{\alpha})}\right]$ and standard deviation $\sigma_{\pi}\left[\Sigma_{k^{\alpha},\pi(q^{\alpha})}\right]$ of the entropy distribution is thus calculated. The indicator function $\Theta_{k^{\alpha},q^{\alpha}}$ measures the difference between $\Sigma_{k^{\alpha},q^{\alpha}}$ and $E_{\pi}\left[\Sigma_{k^{\alpha},\pi(q^{\alpha})}\right]$ in units of $\sigma_{\pi}\left[\Sigma_{k^{\alpha},\pi(q^{\alpha})}\right]$. Panel c): Given a second layer $\beta$, $\widetilde\Theta_{\alpha,\beta}$ characterizes the information about the structure in layer $\alpha$, carried by the characteristics of nodes in layer $\beta$. In order to define a symmetric indicator function of the similarity between the layers $\alpha$ and $\beta$ we define the indicator $\widetilde\Theta^S_{\alpha,\beta}$ that symmetrises the indicator function  $\widetilde\Theta_{\alpha,\beta}$. }
  \label{fig:method}
\end{figure}

We can symmetrize the indicator function  $\widetilde{\Theta}$   by defining
\begin{equation} 
\widetilde{\Theta}_{\alpha,\beta}^{S}=\frac{\widetilde{\Theta}_{\alpha,\beta}+\widetilde{\Theta}_{\beta,\alpha}}{2}.
\end{equation}
which quantify how similar layer $\alpha$ and layer $\beta$ are with respect to their community structure. In Figure $\ref{fig:method}$ we sketch the main aspects of the method used to construct the similarity indicator $\widetilde{\Theta}_{\alpha,\beta}^{S}$. 

Given a multiplex network is it possible to evaluate  the entire symmetric matrix $\widetilde{\Theta}^S$ describing the similarity between any two  layers with respect to their mesoscopic structures $(k^{\alpha},q^{\alpha})$.
This matrix can then be used to construct a network of network formed by the layers of the multiplex dataset. In order to do this we  define a dissimilarity $d_{\alpha,\beta}$ between the layers $\alpha$ and $\beta$ given by 
\begin{equation}
d_{\alpha,\beta}=\left|1-\left|\widetilde\Theta^{S}_{\alpha,\beta}\right|\right|.
\label{d}
\end{equation}
defined for every pair of layers $\alpha,\beta$ with $\alpha\neq \beta$.
This matrix describe the  dissimilarity existing between the community structure of any two pair of distinct layers of the multiplex networks and can be used to draw a network of networks between the layers.

Moreover the dissimilarity (distance) matrix  of elements $d_{\alpha,\beta}$ given by Eq. $(\ref{d})$,  can be further analyzed  performing the average linkage clustering \cite{clust1,clust2,clust3}. This method allows us to subsequently cluster together the layers of the multiplex according to their relative dissimilarities, given that once a generic cluster $C_1$ is formed, the distance $d_{c}(C_1,C_2)$ between $C_1$ and another cluster $C_2$ is defined as the average distance between all pairs of layers in the two clusters:   
\begin{equation}
d_{c}(C_1,C_2)=\frac{1}{{\cal N}(C_1){\cal N}(C_2)}\sum_{\alpha \in C_1}\sum_{\beta\in C_2}d_{\alpha,\beta}
\end{equation}
where ${\cal N}(C_i)$ indicates the number of layers in cluster $C_i$.
From this analysis we can define a dendrogram between the layers of the network of networks { revealing if} the layers appear to be divided into different communities or nested one on the other.

In the  Section\ref{Aarhus} we will use the $\widetilde{\Theta}^S$ to analyze in the Aarhus Multiplex Social Network which type of social interactions determine similar mesoscopic features in the corresponding layer structures.

\section{Case studies}
\label{iii}

\subsection{Multiplex PageRank Analysis of the Physical Review E Citation-Collaboration Network}

Scientific collaboration networks describe how scientists team up in order to generate new scientific results and discoveries. Collaborations networks have been studied widely, in order to understand the emergence of new scientific topics and  the community structure underlying the scientific endeavor \cite{newman2001PNAS,newman2001PRE,lee2010}. \\
Collaboration networks can also be used to assess the centrality of scientists. In fact it is often the case that the most important scientists collaborate with many others, acquiring a large centrality in the collaboration network.
The most famous example is probably the one of Paul Erd\"os  (1913-1996), that collaborated with 511 scientists during his very intense career. His important role in the collaboration network has suggested the formulation of the Erd\"os number, assigned to each scientists and given by the distance  from  Paul Erd\"os in the collaboration network.
Erd\"os is not a unique case in scientific collaboration networks, and the  influential scientists in   disciplines different from mathematics (such as biology or experimental  physics) where influential papers are often the results of larger teams of scientists, are usually very central in the collaboration network.

The centrality of scientists is nevertheless more often measured in terms of their number of citations \cite{redner98}, or in terms of other  measures associated to  the impact of their papers  evaluated in  terms 
of citations (the Hirsch Number \cite{Hirsh}, the  i-10, the newly proposed $o$ indicator \cite{Dorogovtsev}). The determination of the success of papers and scientists is the object of increasing attention and has given rise to the emerging field of the Science of Success \cite{Roberta}.

Here we want to find a centrality measure that takes into account both the centrality of a scientist in terms of his/her position in the collaboration networks, and his/her centrality in the citation network formed by scientists citing each other. In this way we want to capture both the impact of the work of single scientists and their social role in the scientific collaboration networks as catalysts for aggregating teams making significant discoveries.
\begin{figure}
\centering
\includegraphics[width= 0.9\columnwidth]{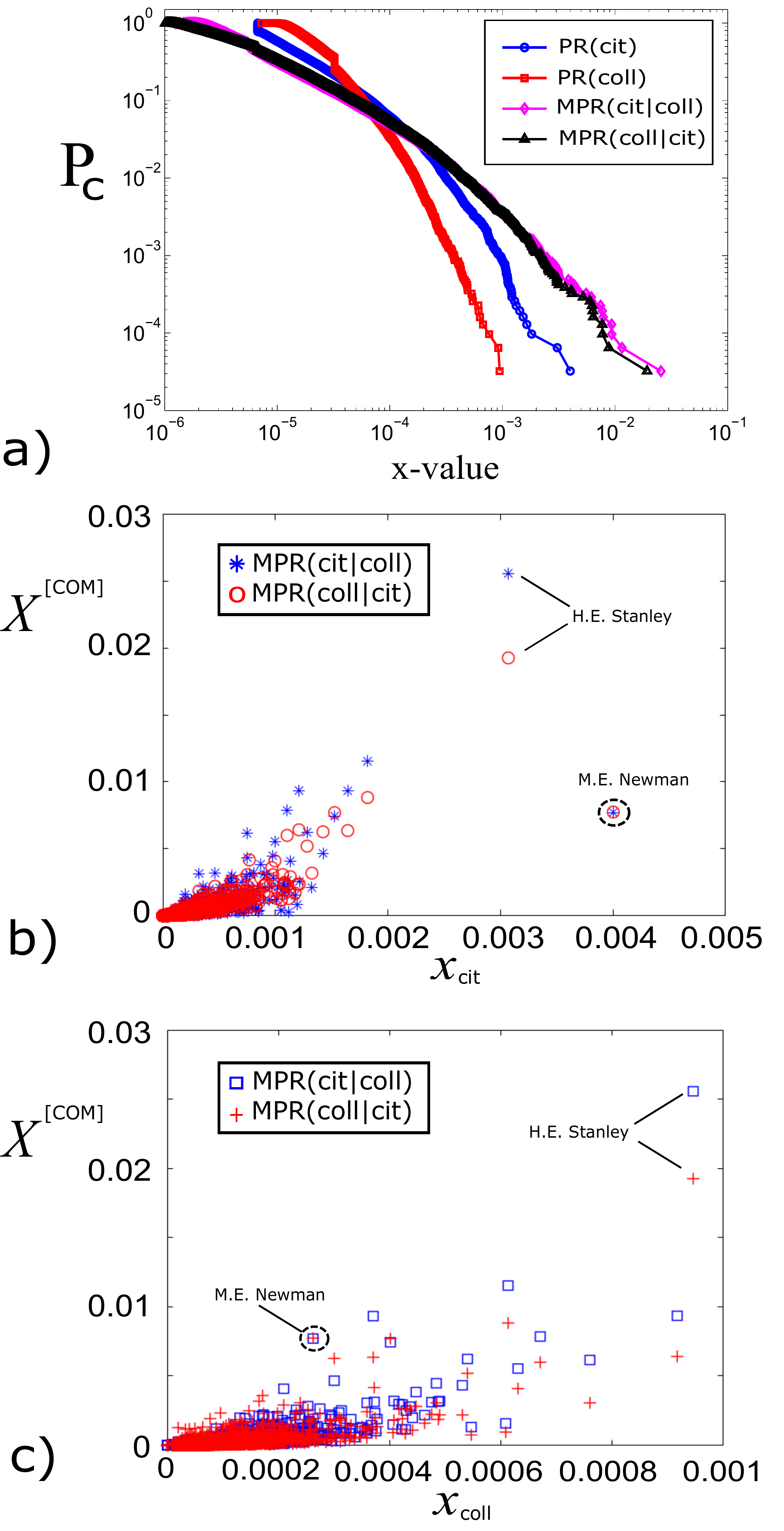}
\caption{Panel a): The cumulative PageRank distribution $P_c(x)$ is shown respectively for the PageRank of the citation PR(cit) and collaboration PR(coll) networks, and for the Combined Multiplex PageRank of the citation network biased by the PageRank on the collaboration network MPR(cit$|$coll) and the one of the collaboration network biased by the PageRank of the citations network MPR(coll$|$cit). All distributions show fat tails but the fat tails are broader for the Multiplex PageRank. Panel b)-c): The Combined Multiplex PageRank MPR(cit$|$coll) and MPR(coll$|$cit) are plotted versus the PageRank PR(cit) and PR(coll). Top cited scientists that are central in the collaboration networks (ex. H. E. Stanley) acquire stronger Multiplex Centrality MPR(cit$|$coll). Scientists that work in small teams or alone but are top cited scientists (ex. M. E. Newman) acquire an increasing relevance in the collaboration network as indicated by their MPR(coll$|$cit). }
\label{fig:MP1}
\end{figure}

To this end we constructed a citation-collaboration multiplex formed by the authors of the Physical Review E (PRE) journal \cite{menichetti2014weighted} . The dataset includes all the papers published on PRE from 1993 to  2009.
Among the papers published in PRE  we focused our study only on those containing a number of authors less or equal to ten. This excludes most of the experimental high-energy collaborations that are typically characterized by a number of authors of a different order of magnitude. We decided to place such a cut-off to the maximum number of
authors allowed per paper to avoid biases due to very large publications. Given the cut-off, our study thus becomes
limited to 35,766 PRE articles and 35,205 PRE authors.
 The first layer characterizes the collaboration network. It is a weighted undirected network with adjacency matrix with elements $a_{ij}^{[1]}>0$ for every pair of authors that have co-authored at least a single common paper.
The weight of the links $a_{ij}^{[1]}=w_{ij}^{[1]}$ are calculated according to the expression
\bea
a_{ij}=\sum_{p\in I}\frac{\xi_{i}^p \xi_j^p}{n_p-1},
\eea
where $I$ indicates the set of all papers $p$ in the dataset, $n_p$ indicates the number of authors of the paper $p$ and $\xi_i^p$ is an indicator function taking value $\xi_i^p=1$ if the scientist $i$ is a co-author of paper $p$.  
The second layer of citation network is directed and weighted with adjacency matrix $a_{ij}^{[2]}=w_{ij}^{[2]}$ indicating how many times the scientist $j$ has cited the scientist $i$.
 \begin{figure}
\includegraphics[width= 0.85\columnwidth]{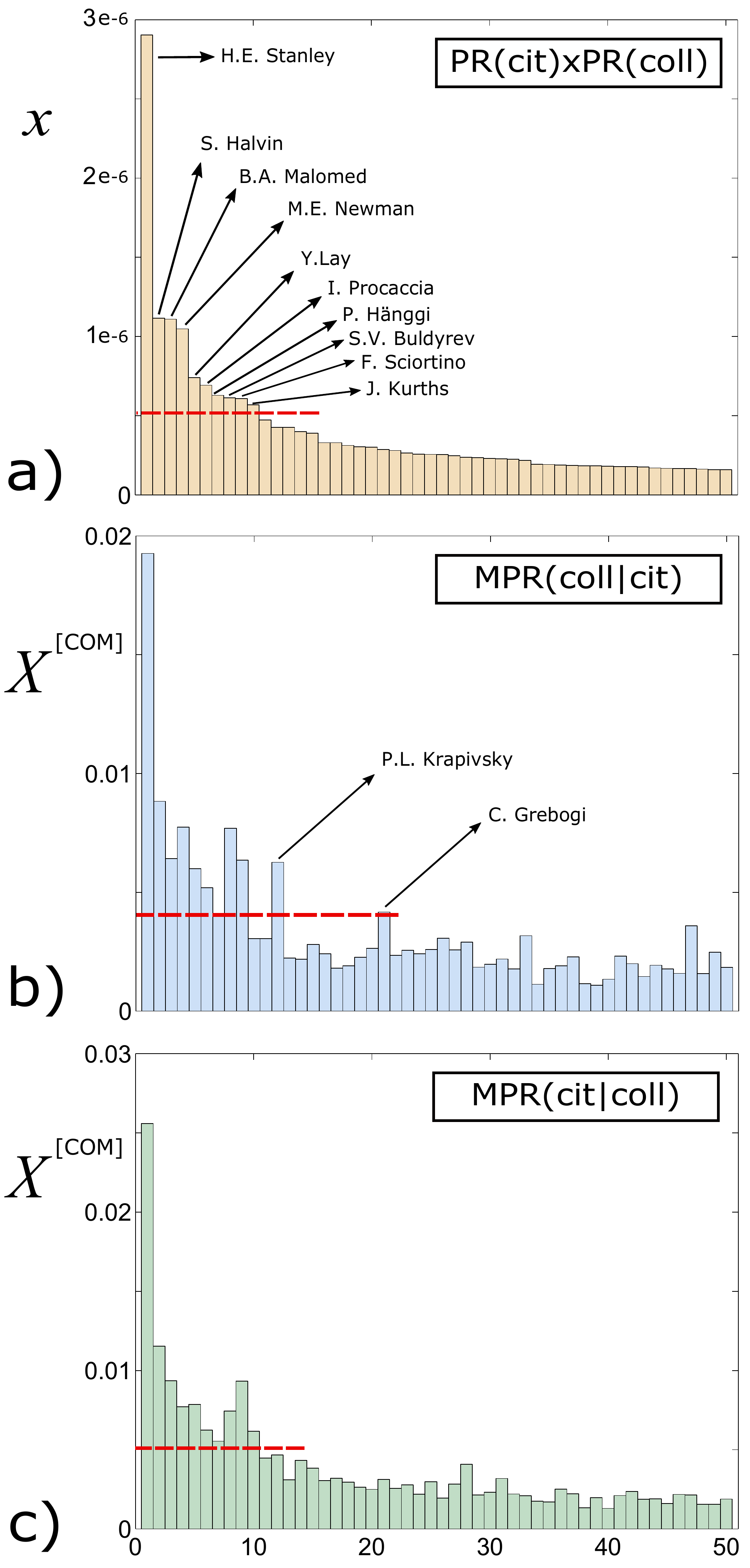} 
  \caption{The centrality  of the top fifty ranked authors according to the product of the two PageRanks PR(coll) and PR(cit) is shown in panel a).
  In the other two panels the Multiplex PageRank 
  PR(coll$|$cit) (panel b)) and to the Multiplex PageRankPR(cit$|$coll) (panel c)) is shown for the same set of individuals considered in panel a). In all the panels the order of the scientist on the x-axis  follows the ranking  according to the product of   PR(coll) and PR(cit). The dotted red line allows to individuate the the top ten ranked authors, whose centrality values fall above the line.}
  \label{fig:PRErank}
\end{figure}

\begin{table*}
\centering
\caption{The top thirty ranked authors according  respectively to  the  PageRank on citations (PR(cit)),  the Combined Multiplex PageRank on citations given collaborations (MPR(cit$|$coll)), the PageRank on collaborations (PR(coll)),the Combined Multiplex PageRank on collaborations given citations (MPR(coll$|$cit)) and the proxy given by the product of PR(coll) and PR(cit). Here each layer is weighted.}
\label{tab:1}
\begin{ruledtabular}
\begin{tabular}{cccccc}
\hline
\mbox{Rank}  &\mbox{PR(cit)} & \mbox{MPR(cit$|$coll)} &\mbox{PR(coll)} & \mbox{MPR(coll$|$cit)} & \mbox{PR(coll)$\times$PR (cit)}\\ \hline
1  &     M. E. Newman	  \ \ \ \ 	 &  H. E. Stanley	\ \ \ \ \ \ \ 	&  H. E. Stanley	\ \ \ \ 	&  H. E. Stanley  \ \ \ \ 	&  H. E. Stanley
\\
2  &     H. E. Stanley	  	 &  S. Havlin		    &  B. A. Malomed		&  S. Havlin             &  S. Havlin
\\
3  &     S. Havlin	  	     &	B. A. Malomed		&  J. Kurths		    &  M. E. Newman          &  B. A. Malomed
\\
4  &     F. Sciortino	  	 &	F. Sciortino		&  Y. Lai		        &  S. V. Buldyrev        &  M. E. Newman
\\
5  &     S. V. Buldyrev	  	 &	Y. Lai		        &  P. H\"{a}nggi		&  B. A. Malomed         &  Y. Lai
\\
6  &     P. L. Krapivsky	 &  M. E. Newman		&  S. Havlin		    &  F. Sciortino          &  I. Procaccia
\\
7  &     A. Vespignani	  	 &	S. V. Buldyrev	    &  J. Zhang		        &  P. L. Krapivsky       &  P. H\"{a}nggi
\\
8  &     I. Procaccia  	     &  I. Procaccia		&  J. Wang		        &  Y. Lai                &	S. V. Buldyrev
\\
9  &     F. H. Stillinger    & J. Kurths 		    &  I. Procaccia		    &  I. Procaccia          &  F. Sciortino
\\
10 &     B. A. Malomed		 & P. H\"{a}nggi 		&  S. Chen		        &  C. Grebogi            & J. Kurths
\\
11 &     S. Luding           & P. L. Krapivsky		&  B. Hu		        &  P. H\"{a}nggi         &	E. Ott
\\
12 &   S. H. Strogatz        &	E. Ott				&  P. G. Kevrekidis	    & E. Ben-Naim            & P. L. Krapivsky
\\
13 &   P. Tartaglia          &	S. Chen				&  E. Ott		        & A. Vespignani          &  H. J. Herrmann
\\
14 &   D. J. Watts           &	P. Tartaglia		&  H. J. Herrmann 		& A. Blumen              &	S. Chen
\\
15 &   W. G\"{o}tze         &	Y. S. Kivshar		&  W. Wang		        & E. Ott                 &	Y. S. Kivshar
\\
16 &   A. J. Bray            &  S. Dietrich		    &  Y. S. Kivshar		& J. Kurths              &	S. Torquato
\\
17 &   Y. Lai                &	B. Hu				&  S. Kim 	            & P. Tartaglia           & S. Dietrich
\\
18 &   A. R. Bulsara         &	P. G. Kevrekidis	&  S. {\v Z}umer		& Y. S. Kivshar          &	H. L\"{o}wen
\\
19 &   R. Pastor-Satorras    &	C. Grebogi		    &  H. Chen		        & H. Chen                &	A. J. Bray
\\
20 &   S. R. Nagel           &	H. J. Herrmann		&  Y. Chen		        & A. R. Bishop           &  H. Chen
\\
21 &   A. Blumen             &	S. Torquato         & H. L\"{o}wen	        & L. S. Tsimring         &	C. Grebogi
\\
22 &   M. Fuchs			     &  A. R. Bishop        & G. Barbero		    & S. {\v Z}umer          & F. H. Stillinger
\\
23 &   R. T. Farouki         &	H. L\"{o}wen		& A. R. Bishop		    & R. Pastor-Satorras     & S. {\v Z}umer
\\
24 &   S. Hamaguchi          &	L. S. Tsimring		& S. Dietrich		    & S. Torquato            & A. R. Bulsara
\\
25 &   H. D. Abarbanel		 &  S. {\v Z}umer 	    & W. Wang		        & A. R. Bulsara          &  A. R. Bishop
\\
26 &   W. Kob                &	A. J. Bray		    & S. V. Buldyrev		& F. H. Stillinger       &   A. Blumen
\\
27 &   P. H\"{a}nggi        &	F. H. Stillinger	& G. Hu		            & S. Luding              &	L. S. Tsimring
\\
28 &   L. S. Tsimring        &	F. Lederer		    & C. Hu		            & S. Redner              & P. Tartaglia
\\
29 &   E. Ott                &	H. Chen 	        & K. Binder		        & S. R. Nagel            &   H. D. Abarbanel
\\
30 &   E. Ben-Naim           &	S. N. Majumdar	    & C. Grebogi		    & A. J. Bray             & K. Binder       
\end{tabular}
\end{ruledtabular}
\end{table*}

We first ranked scientists using the standard PageRank both on the layer of collaboration PR(coll) and on the layer of citation PR(cit). Then we performed a Combined Multiplex PageRank  on the network of citation given the PageRank on the network of collaboration MPR(cit$|$coll) and a Combined Multiplex PageRank on the network of collaboration given the  PageRank on the network of citation MPR(coll$|$cit).  We took always  into consideration the weights  of the multiplex network. The damping factors $\mu_A$ and $\mu_B$ are taken to be both equal to 0.85, i.e. $\mu_A=\mu_B=0.85$.
The distribution of the PageRank and the Multiplex PageRank are all broadly distributed allowing for a well defined ranking of the top scientists (see Figure $\ref{fig:MP1}a$).
The effect of the Multiplex PageRank on the ranking of the authors can be very significant since the rankings in the citation and collaboration network can differ significantly (see Figure $\ref{fig:MP1}b$ and $\ref{fig:MP1}c$).
In particular the MultiplexPageRank (coll$|$cit) boost the centrality of highly cited scientists in the collaboration network, and the Multiplex PageRank (cit$|$coll) boost the centrality in the citation network of the top ranked authors in the collaboration network.
{It turns out that a good proxy for the Multiplex PageRank of the top scientists in the MPR(coll$|$cit) and MPR(cit$|$coll) is the product of the two PageRanks PR(coll) and PR(cit).}
In Figure \ref{fig:PRErank} we report the rankings of PRE authors according to the Multiplex PageRank PR(coll$|$cit), PR(cit$|$coll) and the product of the two PageRanks PR(coll) and PR(cit).  
In Table \ref{tab:1} we report the first thirty authors ranked respectively with the  PageRank on citations,  PageRank on collaborations, the Combined Multiplex PageRank on citations given collaborations and Combined Multiplex PageRank on collaborations given citations and the proxy given by the product of PR(coll) and PR(cit). 

To assess the similarity of the rankings we consider the set of the $n_{100}$ authors present in the union of the MPR(coll$|$cit), MPR(cit$|$coll) and PR(cit)xPR(coll)  and we measure  the Kendall tau correlation coefficient defined as    
\begin{equation}
\tau_K=\frac{\tau_1 - \tau_0}{n_{100}(n_{100}-1)/2}
\end{equation}
where $\tau_1$ is the number of pairs of authors whose order in the two different rank lists considered is concordant while $\tau_0$ is the number of pairs of authors whose order in the two different rank lists considered is discordant.  
We get that the Kendall tau correlation between MPR(cit$|$coll) and PR(cit)xPR(coll)  is 0.76, while the Kendall tau correlation between MPR(coll$|$cit) and PR(cit)xPR(coll)  is 0.56.

This analysis shows that the centrality of an individual on a multiplex network is  strongly affected by synergistic effects due to the correlated structure of   its different layers.
The Multiplex PageRank is able to capture exactly these effects, as we have shown here in the case of the Multiplex PageRank on the collaboration-citation network.
Measuring the centrality of an author only according to his citation network or his collaboration network provides only a partial account of the centrality of the author.
In fact the centrality of a scientists is strongly affected both by its citation record and its network of collaboration.
The Multiplex PageRank instead fully accounts for both layers and evaluates the role of an author as a catalyst of scientific innovation, i.e. at the same time publishing highly cited papers and actively promoting scientific collaborations.

\subsection{Multiplex PageRank Analysis of Noordin Terrorist Network}
\label{noordin}

 In this section we analyze the so called Noordin Top Terrorist Network \cite{nancy2011} using the Multiplex PageRank previously defined and we compare our results with the results obtained using the standard PageRank. Noordin Mohammed Top was an Indonesian terrorist who built a personal terrorist group and was responsible of several bombing attacks in Malaysia between 2003 and 2005. We consider here, with an approach similar to \cite{Battiston_measures}, the data subset consisting of 79 terrorists for three different types of relations: 1) the network of the trust, which considers friendship, family connections and school or religious affiliations between the terrorists; 2) the network of the communication, which considers messages exchanged between the terrorists; 3) the network of the operations, which considers operational relations like participate to the same terrorist operation or having joined the same training camp.

\begin{figure}
\centering
\includegraphics[width= 9 cm]{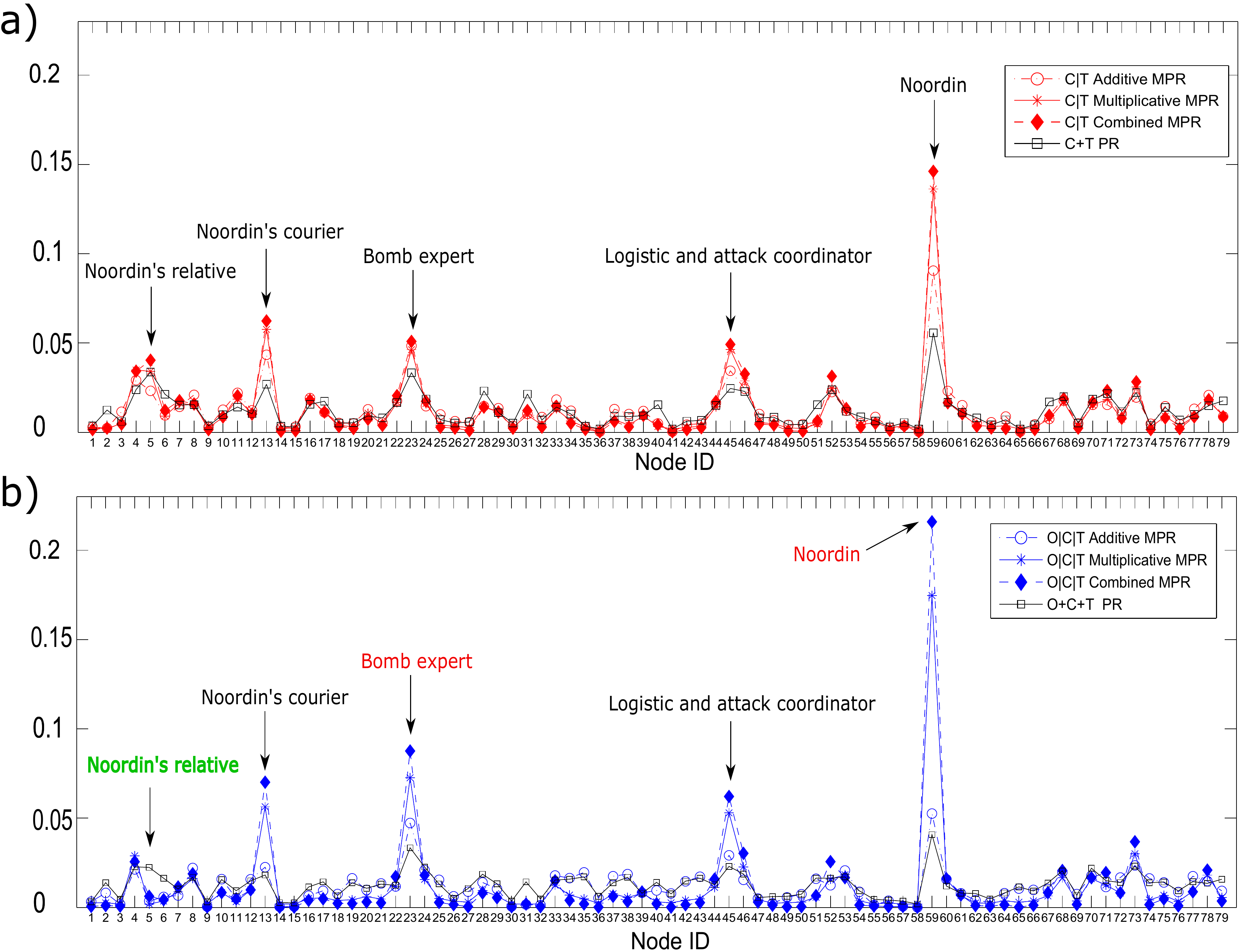}
\caption{
Multiplex PageRank analysis of the Multiplex Noordin Terrorist Network consisting of 79 terrorists and three different levels of interaction: the network of the trust(T),  
the network of the communication (C)and the network of the operations(O). Panel a): Additive (circles), Multiplicative (stars) and Combined (diamonds) Multiplex PageRanks of the terrorists in the communication network, given the standard PageRank in the network of trust (C$|$T), are compared with a standard PageRank (squares) of nodes in the network resulting from the aggregation of the layer of trust and the layer of communication (C+T). The Multiplex PageRank clearly indicates people who can have a relevant role in the organization (nodes indicated by arrows). Panel b): Additive (circles), Multiplicative (stars) and Combined (diamonds) Multiplex PageRanks of the terrorists in the operational network, given the respective Multiplex PageRanks in the network of communication (O$|$C$|$T), are compared with a standard PageRank (squares) of nodes in the network resulting from aggregating the layer of operation, communication and trust (O+C+T). Compared with the two level Multiplex PageRank the three level Multiplex PageRank is able to correctly distinguish people who effectively had a prominent role in the communications and in the operations (names in red), people who had a prominent role only at the level of communication (black names) and people with a minor role in the organization but highly ranked in the communication network (green names).}
\label{fig:2}
\end{figure}

\begin{figure*}
\centering
\includegraphics[width= 13cm]{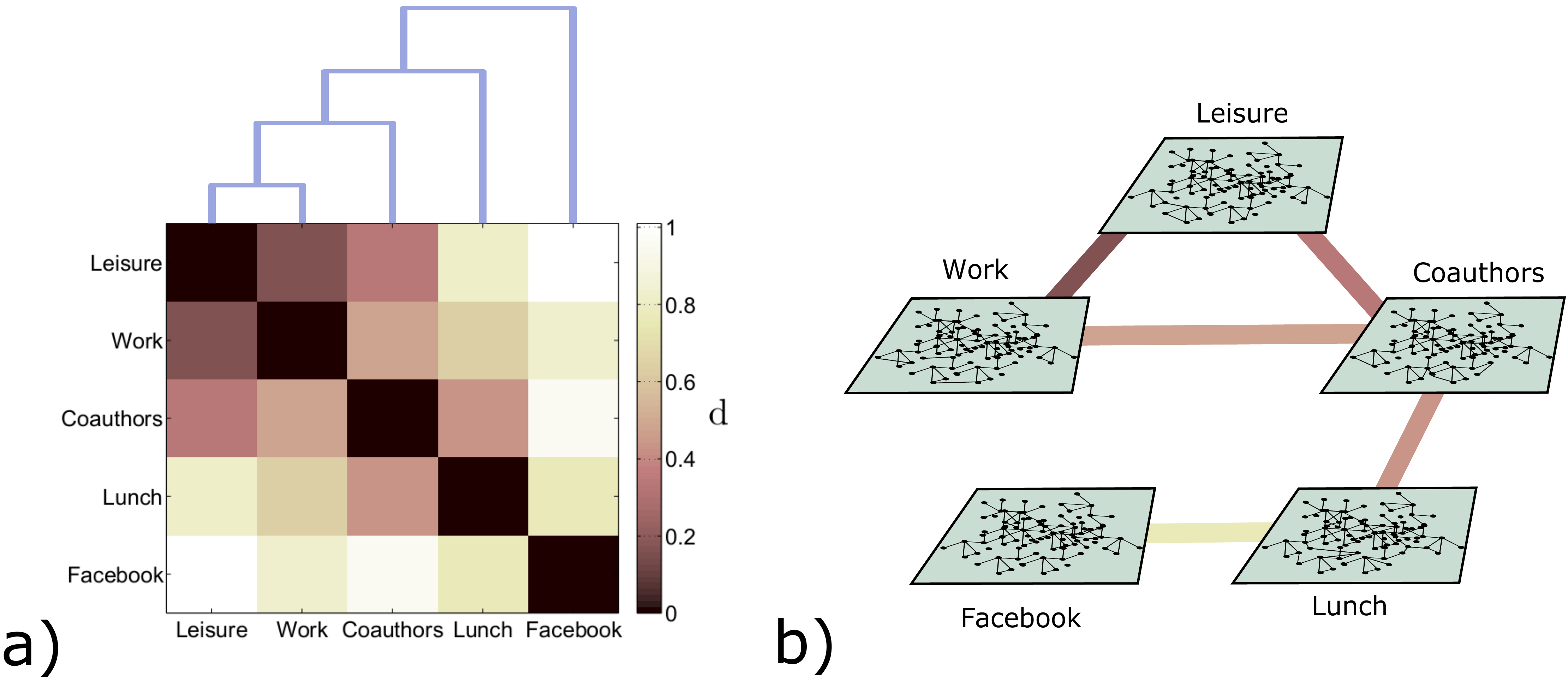}
\caption{Panel a): The dissimilarity  matrix of elements $d_{\alpha,\beta}$  is shown for the Aarhus Multiplex Social Network where the 5 layers describe different types of interactions between the employees of a Computer Science department. Pairs of layers for which the value of $d_{\alpha,\beta}$ is smaller present a similar mesoscopic structure. The dendrogram between the layers extracted from the matrix $d_{\alpha,\beta}$, shown on top, reveals a hierarchical  structure between the different types of social interactions (layers). Panel b): The network between the layers is shown. Here the nodes corresponds to the layers of the multiplex and the weighted links correspond to the values $d_{\alpha,\beta}$ of dissimilarity. Only links with dissimilarity $d_{\alpha,\beta}$ below a threshold  are shown. This threshold  is the minimum dissimilarity value for which the graph is all connected.}     
\label{fig:3}
\end{figure*}

 This multiplex terrorist network is an interesting case where to apply the Multiplex PageRank. Indeed it is reasonable to assume that terrorists prominence in the network of trust is likely to contribute to their prominence in the communication network. Similarly the terrorist rank in in the operational network is reasonably affected by their rank in the communication network.
 
We first measured the standard PageRank of the terrorists in the network of trust (the damping parameters are taken to be $\mu_A=\mu_B=0.85$). Then we measured Additive, Multiplicative and Combined Multiplex PageRank for the terrorists in the communication network and we compared the three ranking obtained with the one obtained by measuring a standard PageRank on the network resulting from the aggregation of the network of trust and the network of the communication. Results are shown in Fig \ref{fig:2} panel a). 
Interestingly all the versions of the Multiplex PageRank on the collaboration network of the communication given the network of the trust (C$|$T) seems to distinguish much more clearly the people with important roles in the organization than the standard PageRank on the aggregated network (C+T).
Among the first five ranked nodes we find four of the terrorists who are actually relevant in the organization: Noordin, a bomb expert, the coordinator of the logistic and of the attacks, the Noordin's courier (black arrows), and one individual, a relative of Noordin, who can result prominent because he is likely to communicate with the leader of the organizations for several different reasons.     

  As a following step in the analysis we measured Additive, Multiplicative and Combined Multiplex PageRank in the operational network using respectively the Additive, Multiplicative and Combined Multiplex PageRank measured previously in the communication network and we measured the obtained ranking of operational given communication given trust (O$|$C$|$T) with the standard PageRank on the network resulting from the aggregation of the three layers (O+C+T), Figure \ref{fig:2} panel b). Again the Multiplex PageRank stresses which people play a crucial role in the organization. We observe that the rank of Noordin and the rank of the bomb expert is increased respect to the previously measured rank, denoting the importance of these two nodes also in the operations, while the importance of the Noordin's courier and the logistic coordinator has not changed, which reflects that they don't play a relevant role in terrorist operations. Surprisingly the rank of the Noordin's relative is dropped down, suggesting that the role of this person in the organization was not really prominent (indeed he didn't have any particular operational skill) and his high rank as a node is probably an effect of being connected with Noordin node in the link of communications.

\subsection{Detecting Mesoscopic Structural Similarities between the Layers of the Multiplex Aarhus Social Network}
\label{Aarhus}
 In this section we analyse the mesocale organization of the multiplex network  called the CS-Aarhus Collaboration Network \cite{magnani2013} using the $\widetilde\Theta^{S}$ measure denied in Sec. \ref{thetasec}. The Aarhus  social network consists of 5 kinds of online and offline relationships (Facebook, Leisure, Work, Co-authorship, Lunch) between the 61 employees of Computer Science department at Aarhus. Our analysis is here used to construct a network between the layer of the multiplex revealing similarity between their mesoscale structure and to test if the mesocale structure of these layers reveal some form of hierarchy between the layers.

The similarity matrix $\widetilde\Theta^{S}$ is constructed using Infomap algorithm \cite{rosvall2007information} for extracting the community structure of the layers   and averaging over $350$ random permutations of the node community labels. 
This similarity values $\widetilde\Theta^{S}_{\alpha,\beta}$ between the layers of the multiplex, can be regarded as the weights of the edges of a fully-connected network (whose nodes represent the 5 different layers of social interactions) which reveals how community of people are organized across different levels of their social life.

The dendrogram resulting from the hierarchical clustering analysis of the matrix $d$ (shown in Figure $\ref{fig:3}a$ reveals that there is an apparent hierarchy between the layers . The Leisure layer and the Work layer are  the most  close to each other  revealing the most significant overlap of communities, while the Facebook layer result to be the less informative network with respect to the division into communities in the entire multiplex dataset.

Given any multiplex network is thus possible, using this general method based on the $\widetilde\Theta^{S}$ indicator, to extract a network between the layers which describes how the mesoscopic structures $(k^{\alpha},q^{\alpha})$ of the layers correlate, and to reduce the information of this network of networks by finding an optimal partition of the layers into cluster at any desired level of correlation. In particular in Figure $\ref{fig:3}b$ the network between the layers is constructed by considering as weighted edges in between the layers only those values of dissimilarity $d_{\alpha,\beta}$ below the minimum threshold which guarantees that the graph is fully connected.

\section{Conclusions}
\label{iv}

Large network datasets as different as social networks, transportation networks or cellular and brain networks are described as multiplex structures where nodes are connected by different types of interactions.
Here we have discussed two measures, namely the Multiplex PageRank and the $\widetilde\Theta^{S}$ indicator function and we have shown that they are useful tools to extract information from multiplex networks. In particular we have applied the Multiplex PageRank to the analysis of the Physical Review E Citation-Collaboration Network, suggesting a new rank of  the authors based  at the same time on their citation impact and on their collaboration activity.   We have also applied the Multiplex PageRank to  the Noordin Terrorist Network, revealing which terrorists played a central role in the organization both at the level of the logistic and of the operations.     
Finally we have described a methodology which makes use of the $\widetilde\Theta^{S}$ indicator to extract the network between the layers of the CS-Aarhus Social Network and allowing us to identify how the organization in communities of  social networks correlates across different social activities.


\begin{thebibliography}{}

\expandafter\ifx\csname natexlab\endcsname\relax\def\natexlab#1{#1}\fi
\expandafter\ifx\csname bibnamefont\endcsname\relax
  \def\bibnamefont#1{#1}\fi
\expandafter\ifx\csname bibfnamefont\endcsname\relax
  \def\bibfnamefont#1{#1}\fi
\expandafter\ifx\csname citenamefont\endcsname\relax
  \def\citenamefont#1{#1}\fi
\expandafter\ifx\csname url\endcsname\relax
  \def\url#1{\texttt{#1}}\fi
\expandafter\ifx\csname urlprefix\endcsname\relax\def\urlprefix{URL }\fi
\providecommand{\bibinfo}[2]{#2}
\providecommand{\eprint}[2][]{\url{#2}}



\bibitem[{\citenamefont{Boccaletti et~al.}(2014)}]{PhysicsReports}
\bibinfo{author}{\bibfnamefont{S.}~\bibnamefont{Boccaletti}},
  \bibinfo{author}{\bibfnamefont{G.}~\bibnamefont{Bianconi}},
  \bibinfo{author}{\bibfnamefont{R.}~\bibnamefont{Criado}},
  \bibinfo{author}{\bibfnamefont{C.}~\bibnamefont{Del~Genio}},
  \bibinfo{author}{\bibfnamefont{J.}~\bibnamefont{G{\'o}mez-Garde{\~n}es}},
  \bibinfo{author}{\bibfnamefont{M.}~\bibnamefont{Romance}},
  \bibinfo{author}{\bibfnamefont{I.}~\bibnamefont{Sendina-Nadal}},
  \bibinfo{author}{\bibfnamefont{Z.}~\bibnamefont{Wang}}, \bibnamefont{and}
  \bibinfo{author}{\bibfnamefont{M.}~\bibnamefont{Zanin}},
  \bibinfo{journal}{Physics Reports} \textbf{\bibinfo{volume}{544}},
  \bibinfo{pages}{1} (\bibinfo{year}{2014}).

\bibitem[{\citenamefont{Kivel{\"a} et~al.}(2014)}]{Kivela}
\bibinfo{author}{\bibfnamefont{M.}~\bibnamefont{Kivel{\"a}}},
  \bibinfo{author}{\bibfnamefont{A.}~\bibnamefont{Arenas}},
  \bibinfo{author}{\bibfnamefont{M.}~\bibnamefont{Barthelemy}},
  \bibinfo{author}{\bibfnamefont{J.~P.} \bibnamefont{Gleeson}},
  \bibinfo{author}{\bibfnamefont{Y.}~\bibnamefont{Moreno}}, \bibnamefont{and}
  \bibinfo{author}{\bibfnamefont{M.~A.} \bibnamefont{Porter}},
  \bibinfo{journal}{Journal of Complex Networks} \textbf{\bibinfo{volume}{2}},
  \bibinfo{pages}{203} (\bibinfo{year}{2014}).
  
  
  
\bibitem[{\citenamefont{Bianconi}(2015)}]{Interdisciplinary}
\bibinfo{author}\bibfnamefont{G.}~\bibnamefont{Bianconi}  \bibinfo{journal}{EPL (Europhysics Letters)} \textbf{\bibinfo{volume}{111}},
  \bibinfo{pages}{56001} (\bibinfo{year}{2015}). 
  

\bibitem[{\citenamefont{Fienberg et~al.}(2008)}]{Wasserman}
S. E. Fienberg,  M. M. Meyer and S. S. Wasserman, Journal of the American Statistical association \textbf{80}, 389 (1985).


\bibitem[{\citenamefont{Cardillo et~al.}(2012)}]{Cardillo}
\bibinfo{author}{\bibfnamefont{A.}~\bibnamefont{Cardillo}},
  \bibinfo{author}{\bibfnamefont{M.}~\bibnamefont{Zanin}},
  \bibinfo{author}{\bibfnamefont{J.}~\bibnamefont{G{\'o}mez-Garde{\~n}es}},
  \bibinfo{author}{\bibfnamefont{M.}~\bibnamefont{Romance}},
  \bibinfo{author}{\bibfnamefont{A.~J.~G.} \bibnamefont{del Amo}},
  \bibnamefont{and}
  \bibinfo{author}{\bibfnamefont{S.}~\bibnamefont{Boccaletti}},
   \bibinfo{journal}{Scientific reports} \textbf{\bibinfo{volume}{3}} (\bibinfo{year}{2013}).




\bibitem[{\citenamefont{Nicosia et al.}(2014)}]{nicosia_correlations}
\bibinfo{author}{\bibfnamefont{V.}~\bibnamefont{Nicosia}} \bibnamefont{and}
  \bibinfo{author}{\bibfnamefont{V.}~\bibnamefont{Latora}},
  \bibinfo{journal}{Phys. Rev. E 89, 032804 }  (\bibinfo{year}{2015}).


\bibitem[{\citenamefont{Menichetti et~al.}(2014)}]{menichetti2014weighted}
\bibinfo{author}{\bibfnamefont{G.}~\bibnamefont{Menichetti}},
  \bibinfo{author}{\bibfnamefont{D.}~\bibnamefont{Remondini}},
  \bibinfo{author}{\bibfnamefont{P.}~\bibnamefont{Panzarasa}},
  \bibinfo{author}{\bibfnamefont{R.~J.} \bibnamefont{Mondrag{\'o}n}},
  \bibnamefont{and} \bibinfo{author}{\bibfnamefont{G.}~\bibnamefont{Bianconi}},
  \bibinfo{journal}{PloS one} \textbf{\bibinfo{volume}{9}},
  \bibinfo{pages}{e97857} (\bibinfo{year}{2014}).



\bibitem[{\citenamefont{Bullmore et al.}(2009)}]{Bullmore}
\bibinfo{author}{\bibfnamefont{E.}~\bibnamefont{Bullmore}} \bibnamefont{and}
  \bibinfo{author}{\bibfnamefont{O.}~\bibnamefont{Sporns}},
  \bibinfo{journal}{Nature Reviews Neuroscience} \textbf{\bibinfo{volume}{10}},
  \bibinfo{pages}{186} (\bibinfo{year}{2009}).


\bibitem[{\citenamefont{Reis et~al.}(2014)}]{Makse}
S. DS. Reis, Y. Hu, A. Babino, J.Ž S. Andrade Jr, S. Canals, M. Sigman, and H. A. Makse, Nature Physics \textbf{10}, 10 (2014).


\bibitem[{\citenamefont{Battiston et~al.}(2014)}]{Battiston_measures}
\bibinfo{author}{\bibfnamefont{F.}~\bibnamefont{Battiston}},
  \bibinfo{author}{\bibfnamefont{V.}~\bibnamefont{Nicosia}}, \bibnamefont{and}
  \bibinfo{author}{\bibfnamefont{V.}~\bibnamefont{Latora}},
  \bibinfo{journal}{Physical Review E} \textbf{\bibinfo{volume}{89}},
  \bibinfo{pages}{032804} (\bibinfo{year}{2014}).



\bibitem[{\citenamefont{Cantini et~al.}(2014)}]{Caselle} L. Cantini, E. Medico, S. Fortunato and M. Caselle, Scientific Reports \textbf{5} (2015).




\bibitem[{\citenamefont{Lee et~al.}(2014)}]{Goh_rev}
K-M. Lee, J. Y. Kim, S. Lee, and K-I. Goh, Networks of networks: The last frontier of complexity, pp. 53-72. Springer International Publishing, 2014.

\bibitem{Cellai2013}
D. Cellai, E. L{\'o}pez, J. Zhou, J. P. Gleeson, and G. Bianconi Physical Review E \textbf{88}, 5 (2013).

\bibitem{Cellai2016}
D. Cellai, S. N. Dorogovtsev, and G. Bianconi, arXiv preprint arXiv:1604.05175 (2016).

\bibitem{Bianconi2014}
G. Bianconi, and S. N. Dorogovtsev, Physical Review E \textbf{89}, 6 (2014): 062814.

\bibitem{synchronization1} L. V. Gambuzza, M. Frasca, and J. {G{\'o}mez-Garde{\~n}es}, EPL (Europhysics Letters) 110, 2 (2015).


\bibitem{synchronization2} X. Zhang, S. Boccaletti, S. Guan, and Z. Liu, Physical review letters \textbf{114}, 3 (2015)

\bibitem{epidemic} C. Granell, S. G{\'o}mez, and A. Arenas, Physical review letters \textbf{111}, 12 (2013)

\bibitem{gametheory1} J. {G{\'o}mez-Garde{\~n}es}, I. Reinares, A. Arenas, and L. M. Flor{\'i}a, Scientific reports \textbf{2} (2012).

\bibitem{gametheory2} J. T. Matamalas, J. Poncela-Casasnovas, S. G{\'o}mez, and A. Arenas, Scientific reports \textbf{5} (2015).

\bibitem{congestion} A. Sol{\'e}-Ribalta, Sergio G{\'o}mez, and A. Arenas, Physical Review Letters \textbf{116}, 10 (2016)

\bibitem{Buldyrev2010} S. V. Buldyrev, R. Parshani, G. Paul, H. E. Stanley, and S. Havlin, Nature \textbf{464}, 7291 (2010).

\bibitem[{\citenamefont{Gomez et~al.}(2013)}]{Diaz}
\bibinfo{author}{\bibfnamefont{S.}~\bibnamefont{Gomez}},\bibinfo{author}{\bibfnamefont{D.}~\bibnamefont{Diaz-Guilera}},\bibinfo{author}{\bibfnamefont{J.}~\bibnamefont{Gomez-Garde\~nes}},\bibinfo{author}{\bibfnamefont{Y.}~\bibnamefont{Moreno}}, \bibnamefont{and} \bibinfo{author}{\bibfnamefont{A.}~\bibnamefont{Arenas}},
  \bibinfo{journal}{Physical Review Letters} \textbf{\bibinfo{volume}{110}},
  \bibinfo{pages}{028701} (\bibinfo{year}{2013}).

\bibitem[{\citenamefont{Radicchi et~al.}(2013)}]{Radicchi} 
\bibinfo{author}{\bibfnamefont{F.}~\bibnamefont{Radicchi}}, \bibnamefont{and} \bibinfo{author}{\bibfnamefont{A.}~\bibnamefont{Arenas}},
  \bibinfo{journal}{Nature Physics} \textbf{\bibinfo{volume}{9}},
  \bibinfo{pages}{717} (\bibinfo{year}{2013}).

\bibitem[{\citenamefont{Szell et~al.}(2010)}]{szell2010multirelational}
\bibinfo{author}{\bibfnamefont{M.}~\bibnamefont{Szell}},
  \bibinfo{author}{\bibfnamefont{R.}~\bibnamefont{Lambiotte}},
  \bibnamefont{and} \bibinfo{author}{\bibfnamefont{S.}~\bibnamefont{Thurner}},
  \bibinfo{journal}{Proceedings of the National Academy of Sciences}
  \textbf{\bibinfo{volume}{107}}, \bibinfo{pages}{13636}
  (\bibinfo{year}{2010}).


\bibitem[{\citenamefont{Bianconi}(2013)}]{PRE}
\bibinfo{author}{\bibfnamefont{G.}~\bibnamefont{Bianconi}},
  \bibinfo{journal}{Physical Review E} \textbf{\bibinfo{volume}{87}},
  \bibinfo{pages}{062806} (\bibinfo{year}{2013}).
 
 
 
\bibitem[{\citenamefont{Min et~al.}(2014)}]{Goh_correlations}
\bibinfo{author}{\bibfnamefont{B.}~\bibnamefont{Min}},
  \bibinfo{author}{\bibfnamefont{S.}~\bibnamefont{Do~Yi}},
  \bibinfo{author}{\bibfnamefont{K.-M.} \bibnamefont{Lee}}, \bibnamefont{and}
  \bibinfo{author}{\bibfnamefont{K.-I.} \bibnamefont{Goh}},
  \bibinfo{journal}{Physical Review E} \textbf{\bibinfo{volume}{89}},
  \bibinfo{pages}{042811} (\bibinfo{year}{2014}).



\bibitem[{\citenamefont{Cellai et al.}(2015)}]{Cellai_activity}
\bibinfo{author}{\bibfnamefont{D.}~\bibnamefont{Cellai}} \bibnamefont{and}
  \bibinfo{author}{\bibfnamefont{G.}~\bibnamefont{Bianconi}},
  \bibinfo{journal}{Physical Review E} \textbf{\bibinfo{volume}{93}},
  \bibinfo{pages}{032302} (\bibinfo{year}{2016}).

\bibitem[{\citenamefont{Iacovacci et~al.}(2015)}]{iacovacci2015} J. Iacovacci, Z. Wu, and G. Bianconi Physical Review E \textbf{92}, 4 (2015) 


\bibitem[{\citenamefont{Battiston et~al.}(2015)}]{battiston2015emergence}
\bibinfo{author}{\bibfnamefont{F.}~\bibnamefont{Battiston}},
  \bibinfo{author}{\bibfnamefont{J.}~\bibnamefont{Iacovacci}},
  \bibinfo{author}{\bibfnamefont{V.}~\bibnamefont{Nicosia}},
  \bibinfo{author}{\bibfnamefont{G.}~\bibnamefont{Bianconi}}, \bibnamefont{and}
  \bibinfo{author}{\bibfnamefont{V.}~\bibnamefont{Latora}},
  \bibinfo{journal}{PlosOne} \textbf{\bibinfo{volume}{11}},
  \bibinfo{pages}{e0147451} (\bibinfo{year}{2016}).
  
\bibitem[{\citenamefont{Sol{\'a} et~al.}(2013)}]{sola2013eigenvector}
\bibinfo{author}{\bibfnamefont{L.}~\bibnamefont{Sol{\'a}}},
  \bibinfo{author}{\bibfnamefont{M.}~\bibnamefont{Romance}},
  \bibinfo{author}{\bibfnamefont{R.}~\bibnamefont{Criado}},
  \bibinfo{author}{\bibfnamefont{J.}~\bibnamefont{Flores}},
  \bibinfo{author}{\bibfnamefont{A.~G.} \bibnamefont{del Amo}},
  \bibnamefont{and}
  \bibinfo{author}{\bibfnamefont{S.}~\bibnamefont{Boccaletti}},
  \bibinfo{journal}{Chaos: An Interdisciplinary Journal of Nonlinear Science}
  \textbf{\bibinfo{volume}{23}}, \bibinfo{pages}{033131}
  (\bibinfo{year}{2013}).


\bibitem[{\citenamefont{Halu et~al.}(2013)}]{MultiplexPageRank}
\bibinfo{author}{\bibfnamefont{A.}~\bibnamefont{Halu}},
  \bibinfo{author}{\bibfnamefont{R.~J.} \bibnamefont{Mondrag{\'o}n}},
  \bibinfo{author}{\bibfnamefont{P.}~\bibnamefont{Panzarasa}},
  \bibnamefont{and} \bibinfo{author}{\bibfnamefont{G.}~\bibnamefont{Bianconi}},
  \bibinfo{journal}{PloS one} \textbf{\bibinfo{volume}{8}},
  \bibinfo{pages}{e78293} (\bibinfo{year}{2013}).


\bibitem[{\citenamefont{De~Domenico
  et~al.}(2015{\natexlab{a}})}]{de2015ranking}
\bibinfo{author}{\bibfnamefont{M.}~\bibnamefont{De~Domenico}},
  \bibinfo{author}{\bibfnamefont{A.}~\bibnamefont{Sol{\'e}-Ribalta}},
  \bibinfo{author}{\bibfnamefont{E.}~\bibnamefont{Omodei}},
  \bibinfo{author}{\bibfnamefont{S.}~\bibnamefont{G{\'o}mez}},
  \bibnamefont{and} \bibinfo{author}{\bibfnamefont{A.}~\bibnamefont{Arenas}},
  \bibinfo{journal}{Nature communications} \textbf{\bibinfo{volume}{6}}
  (\bibinfo{year}{2015}{\natexlab{a}}).


\bibitem[{\citenamefont{Roberts et~al.}(2011)}]{nancy2011} N. Roberts, and Sean F. Everton. 2011. Roberts and Everton Terrorist Data: Noordin Top Terrorist Network (Subset). 




\bibitem[{\citenamefont{Mucha et~al.}(2010)}]{Mucha}
\bibinfo{author}{\bibfnamefont{P.~J.} \bibnamefont{Mucha}},
  \bibinfo{author}{\bibfnamefont{T.}~\bibnamefont{Richardson}},
  \bibinfo{author}{\bibfnamefont{K.}~\bibnamefont{Macon}},
  \bibinfo{author}{\bibfnamefont{M.~A.} \bibnamefont{Porter}},
  \bibnamefont{and} \bibinfo{author}{\bibfnamefont{J.-P.}
  \bibnamefont{Onnela}}, \bibinfo{journal}{Science}
  \textbf{\bibinfo{volume}{328}}, \bibinfo{pages}{876} (\bibinfo{year}{2010}).



\bibitem[{\citenamefont{Valles-Catala et~al.}(2014)}]{Sales}
\bibinfo{author}{\bibfnamefont{T.}~\bibnamefont{Valles-Catala}},
  \bibinfo{author}{\bibfnamefont{F.~A.} \bibnamefont{Massucci}},
  \bibinfo{author}{\bibfnamefont{R.}~\bibnamefont{Guimera}}, \bibnamefont{and}
  \bibinfo{author}{\bibfnamefont{M.}~\bibnamefont{Sales-Pardo}},
  \bibinfo{journal}{arXiv preprint arXiv:1411.1098}  (\bibinfo{year}{2014}).


\bibitem[{\citenamefont{De~Domenico
  et~al.}(2015{\natexlab{b}})}]{de2014identifying}
\bibinfo{author}{\bibfnamefont{M.}~\bibnamefont{De~Domenico}},
  \bibinfo{author}{\bibfnamefont{A.}~\bibnamefont{Lancichinetti}},
  \bibinfo{author}{\bibfnamefont{A.}~\bibnamefont{Arenas}}, \bibnamefont{and}
  \bibinfo{author}{\bibfnamefont{M.}~\bibnamefont{Rosvall}},
  \bibinfo{journal}{Phys. Rev. X} \textbf{\bibinfo{volume}{5}}
  (\bibinfo{year}{2015}{\natexlab{b}}).


\bibitem[{\citenamefont{De Domenico et~al.}(2014c)}]{MuxViz}
M. De Domenico, Manlio, M. A. Porter, and A. Arenas, Journal of Complex Networks (2014): cnu038.


\bibitem[{\citenamefont{Magnani et~al.}(2013)}]{magnani2013} M. Magnani, B. Micenkova and L. Rossi., arXiv preprint arXiv:1303.4986 (2013).


\bibitem[{\citenamefont{Brin}(1998)}]{brin1998anatomy} S. Brin and L. Page, Computer networks and ISDN systems 30 (1998): 107-117.

\bibitem[{\citenamefont{Bianconi}(2008)}]{bianconi2008entropy}
\bibinfo{author}{\bibfnamefont{G.}~\bibnamefont{Bianconi}},
  \bibinfo{journal}{EPL (Europhysics Letters)} \textbf{\bibinfo{volume}{81}},
  \bibinfo{pages}{28005} (\bibinfo{year}{2008}).

\bibitem[{\citenamefont{Bianconi}(2009)}]{bianconi2009entropy}
\bibinfo{author}{\bibfnamefont{G.}~\bibnamefont{Bianconi}},
  \bibinfo{journal}{Physical Review E} \textbf{\bibinfo{volume}{79}},
  \bibinfo{pages}{036114} (\bibinfo{year}{2009}).

\bibitem[{\citenamefont{Peixoto}(2012)}]{peixoto2012entropy}
\bibinfo{author}{\bibfnamefont{T.~P.} \bibnamefont{Peixoto}},
  \bibinfo{journal}{Physical Review E} \textbf{\bibinfo{volume}{85}},
  \bibinfo{pages}{056122} (\bibinfo{year}{2012}).

\bibitem[{\citenamefont{Bianconi et~al.}(2009)}]{bianconi2009assessing}
\bibinfo{author}{\bibfnamefont{G.}~\bibnamefont{Bianconi}},
  \bibinfo{author}{\bibfnamefont{P.}~\bibnamefont{Pin}}, \bibnamefont{and}
  \bibinfo{author}{\bibfnamefont{M.}~\bibnamefont{Marsili}},
  \bibinfo{journal}{Proceedings of the National Academy of Sciences}
  \textbf{\bibinfo{volume}{106}}, \bibinfo{pages}{11433}
  (\bibinfo{year}{2009}).


 
\bibitem[{\citenamefont{Sokal et al.}(1958)}]{clust1}
\bibinfo{author}{\bibfnamefont{R.}~\bibnamefont{Sokal}} \bibnamefont{and}
  \bibinfo{author}{\bibfnamefont{C.}~\bibnamefont{Michener}},
  \bibinfo{journal}{University of Kansas Science Bulletin}
  \textbf{\bibinfo{volume}{38}}, \bibinfo{pages}{1409} (\bibinfo{year}{1958}).

\bibitem[{\citenamefont{Sokal et al.}(1962)}]{clust2}
\bibinfo{author}{\bibfnamefont{R.}~\bibnamefont{Sokal}} \bibnamefont{and}
  \bibinfo{author}{\bibfnamefont{F.~J.} \bibnamefont{Rohlf}},
  \bibinfo{journal}{Taxon} \textbf{\bibinfo{volume}{11}}, \bibinfo{pages}{33}
  (\bibinfo{year}{1962}).

\bibitem[{\citenamefont{Ying et~al.}(2005)}]{clust3}
\bibinfo{author}{\bibfnamefont{Z.}~\bibnamefont{Ying}},
  \bibinfo{author}{\bibfnamefont{G.}~\bibnamefont{Karypis}}, \bibnamefont{and}
  \bibinfo{author}{\bibfnamefont{U.}~\bibnamefont{Fayyad}},
  \bibinfo{journal}{Data mining and knowledge discovery}
  \textbf{\bibinfo{volume}{10}}, \bibinfo{pages}{141} (\bibinfo{year}{2005}).
 


\bibitem[{\citenamefont{Newman}(2001{\natexlab{a}})}]{newman2001PNAS}
\bibinfo{author}{\bibfnamefont{M.~E.~J.} \bibnamefont{Newman}},
  \bibinfo{journal}{Proceedings of the National Academy of Science}
  \textbf{\bibinfo{volume}{4}}, \bibinfo{pages}{404}
  (\bibinfo{year}{2001}{\natexlab{a}}).


\bibitem[{\citenamefont{Newman}(2001{\natexlab{b}})}]{newman2001PRE}
\bibinfo{author}{\bibfnamefont{M.~E.~J.} \bibnamefont{Newman}},
  \bibinfo{journal}{Physical Review E} \textbf{\bibinfo{volume}{64}},
  \bibinfo{pages}{016132} (\bibinfo{year}{2001}{\natexlab{b}}).



\bibitem[{\citenamefont{Lee et~al.}(2010)}]{lee2010}
\bibinfo{author}{\bibfnamefont{D.}~\bibnamefont{Lee}},
  \bibinfo{author}{\bibfnamefont{K.-I.} \bibnamefont{Goh}},
  \bibinfo{author}{\bibfnamefont{B.}~\bibnamefont{Kahng}}, \bibnamefont{and}
  \bibinfo{author}{\bibfnamefont{D.}~\bibnamefont{Kim}},
  \bibinfo{journal}{Physical Review E} \textbf{\bibinfo{volume}{82}},
  \bibinfo{pages}{026112} (\bibinfo{year}{2010}).


\bibitem[{\citenamefont{Redner}(1998)}]{redner98}
\bibinfo{author}{\bibfnamefont{S.}~\bibnamefont{Redner}},
  \bibinfo{journal}{Eur. Phys. J. B} \textbf{\bibinfo{volume}{4}},
  \bibinfo{pages}{131} (\bibinfo{year}{1998}).
  
  
\bibitem[{\citenamefont{Hirsh}(2005)}]{Hirsh}
{J.~E.} Hirsh, {Proceedings of the National Academy of Sciences}
{\bf 102}, {16569-16572} ({2005}).


\bibitem[{\citenamefont{Dorogovtsev et~al.}(2015)}]{Dorogovtsev}
S. N. Dorogovtsev and J. F. F. Mendes, Nature Physics 11, no. 11 (2015): 882-883.

\bibitem[{\citenamefont{Sinatra et~al.}(2015)}]{Roberta}
R. Sinatra, P. Deville, M. Szell, D. Wang, and A-L. Barab\'{a}si, Nature Physics 11, no. 10 (2015): 791-796.


\bibitem[{\citenamefont{Rosvall et al.}(2007)}]{rosvall2007information}
\bibinfo{author}{\bibfnamefont{M.}~\bibnamefont{Rosvall}} \bibnamefont{and}
  \bibinfo{author}{\bibfnamefont{C.~T.} \bibnamefont{Bergstrom}},
  \bibinfo{journal}{Proceedings of the National Academy of Sciences}
  \textbf{\bibinfo{volume}{104}}, \bibinfo{pages}{7327} (\bibinfo{year}{2007}).




\end{thebibliography}
\end{document}